\documentclass[twoside,a4paper,11pt]{article}
\usepackage[latin1]{inputenc}
\usepackage[T1]{fontenc}
\usepackage{amsmath}
\usepackage{amsfonts}
\usepackage{graphicx}
\usepackage{subfigure}
\usepackage{a4wide}
\usepackage{amssymb}
\usepackage{fancyhdr}
\usepackage{mathrsfs}
\def\beq{\begin{eqnarray}}
\def\eeq{\end{eqnarray}}

\def\be{\begin{eqnarray}}
\def\ee{\end{eqnarray}}

\begin{document}

\begin{center}

{\LARGE\bf Oscillations of the F(R) dark energy \\ in the accelerating universe}
\vspace{5mm}

E.~Elizalde$^{a,}$\footnote{E-mail: elizalde@ieec.uab.es, elizalde@math.mit.edu},
S.~D.~Odintsov$^{a,b,c,}$\footnote{E-mail: odintsov@ieec.uab.es},
L.~Sebastiani$^{d,}$\footnote{E-mail: l.sebastiani@science.unitn.it}
 and S.~Zerbini$^{d,}$\footnote{E-mail: zerbini@science.unitn.it}
\vspace{3mm}

{\small

$^a$Consejo Superior de Investigaciones Cient\'{\i}ficas, ICE/CSIC and IEEC
\\
Campus UAB, Facultat de Ci\`{e}ncies, Torre C5-Parell-2a pl, E-08193
Bellaterra (Barcelona) Spain
\\
$^b$Instituci\'{o} Catalana de Recerca i Estudis Avan\c{c}ats (ICREA)
\\
$^c$Tomsk State Pedagogical University
\\
$^d$Dipartimento di Fisica, Universit\`a di Trento
\\
and Istituto Nazionale di Fisica Nucleare, Gruppo Collegato di Trento, Italia

}

\end{center}

\vspace{3mm}

%%%%%%%%%%%%%%%%%%%%%
%  Abstract
%%%%%%%%%%%%%%%%%%%%%
\begin{abstract}

Oscillations of the $F(R)$ dark energy around the phantom divide line, $\omega_{DE}=-1$, both during the matter era and also in the de Sitter epoch are investigated. The analysis during the de Sitter epoch is revisited by expanding the modified equations of motion around the de Sitter solution. Then,  during the matter epoch,  the time dependence of the dark
energy perturbations is  discussed  by using two different local expansions. For high values of the red shift, the matter epoch is a stable point of the theory, giving the possibility to expand the $F(R)$-functions in terms of the dark energy perturbations. In the late-time matter era, the realistic case is considered where dark energy tends to a constant. The results obtained are confirmed by precise numerical computation on a specific model of exponential gravity. A novel and very detailed discussion is provided on the critical points in the matter era and on the relation of the oscillations with possible singularities.

\end{abstract}
%%%%%%%%%%%%%%%%%%%%%

%----------------------------
%PACS
%----------------------------

%\def\thesection{\Roman{section}}
%\def\theequation{\Roman{section}.\arabic{equation}}
%===========================================================================

\section{Introduction}

Modified gravity constitutes a very powerful and natural possibility to unify
the physics of the early time inflation epoch with that of the late-time acceleration
stage, under the frame of a common theory (for a review, see \cite{F(R)}) and this
without the need to introduce any extra fields (scalar, spinor, or other) as dark
components. Among the possible versions of modified gravity there are in fact a number
of them which comply without problems with all the local tests (for recent reviews, see
\cite{cf10,clf09}) and the most recent and accurate large-scale observational data
(recent discussions can be found in \cite{lssh10,ssh11,saw11}). There is, moreover, a serious
expectation that all of the different stages in the universe evolution: inflation, radiation/matter
dominance, the present and future acceleration epoch, will be qualitatively well described in frames
of a specific class of modified gravity \cite{F(R)}.
Furthermore, modified gravities can be reconstructed to give rise to other realistic scenarios
of the universe evolution, such as the ekpyrotic one \cite{kost1} and the so-called Little Rip cosmology \cite{fls11} (for explicit examples of the late, see \cite{beno11}).

The present paper is devoted to the study of $F(R)$ gravity properties in the matter and in the dark energy epochs. We will use the so-called fluid representation of $F(R)$ gravity \cite{cno06}, where the corresponding equations of motion are presented in the standard gravity, FRW form but with the addition of an effective fluid. Furthermore, this $F(R)$-fluid will here contain an explicit dependence on the gravitational terms. Specifically, we will study the oscillations that are generated in the equation of state (EoS) parameter $w_{DE}$ in $F(R)$, around the phantom divide crossing, during the matter dominance era and in the de Sitter epoch (see also \cite{staro11}).  The evolution of the $F(R)$ dark energy in these regimes will be here carefully investigated, both analytically and numerically.

As is known, an important viable condition in modified gravity is the positivity of the second derivative of the cosmological function ($F''(R)>0$) during the matter dominated era. This condition arises from the stability of matter perturbations in the weak-field regime, taking the curvature to be locally constant. When $F''(R)<0$ perturbations grow up and, as a consequence, the theory becomes strongly unstable \cite{Faraoni}. In this paper a detailed analysis of such perturbations will be carried out, in the approach discussed, by considering the dynamical behavior of dark energy during the matter era. Since in the late-time matter era it is not stable, an expansion of the EOM with respect to the matter curvature is not allowed, and it will be necessary to introduce some physical assumptions in order to find correct results. Accurate viable conditions for the stability of the cosmological perturbations during the matter era will be found, in agreement with the constraint $F''(R)>0$. We will see that an oscillatory behavior of the dark energy can occur during the matter era. This phenomenon leads to oscillations of $\omega_{DE}$, whose frequency and amplitude decay with the red-shift until the current accelerated epoch, while they substantially grow up as a consequence of perturbations around the de Sitter solution of the field equations.

It will be proven, in particular, that the dark energy density approaches a stable de Sitter point, either as a positive power function of $z+1$ or with an oscillatory behavior which amplitude decreases as $(z+1)^{3/2}$. In the late-time matter era, $1\leq z \leq 3$, a realistic, exponential gravity theory will be investigated numerically. We will actually obtain the precise time dependence of the dark energy perturbations. The analysis of this dark energy during the matter epoch will be performed using two different local expansions. For high values of the redshift the matter epoch will turn out to be a stable point of the theory and, thus, it will be possible to expand the $F(R)$-functions in terms of the dark energy perturbations. In the late-time matter era, on the other hand, we will consider the realistic case where dark energy tends to a constant. The results obtained will be carefully checked by numerical computation on a specific model of exponential gravity. The corresponding attractors for this exponential model will be identified.

We include a detailed discussion on the critical points in the matter era. In this set up only the perturbation associated with the scalaron will be investigated, anyhow the results obtained may be interesting and useful. They will be expressed both in semi-analytical and in numerical form, and we will prove their consistency with each other. We do not know of any other competing study on this quite non-trivial subject.
As last point, some problems appearing in relation with the big values generated for the dark energy fluctuations, for large values of the red shift, will be discussed, by considering the relation between the oscillations obtained and possible future singularities.

\section{$\omega_{DE}$-oscillations in realistic $F(R)$ models of the DE-universe}

%\paragraph*{}
The action of modified $F(R)$ theories with matter is \cite{F(R)}
\begin{equation}
S=\int d^4x\sqrt{-g}\left[
\frac{F(R)}{2\kappa^2}+\mathcal{L}^{\mathrm{(matter)}}\right]\,,\label{action}
\end{equation}
where $g$ is the determinant of the metric tensor $g_{\mu\nu}$,
$\mathcal{L}^{\mathrm{(matter)}}$ is the matter Lagrangian and $F(R)$ a generic function of the Ricci scalar, $R$.
%When $F(R)=R$, we recover the Hilbert-Einstein action of General Relativity (GR).
Throughout the paper we will use units where $k_{B}=c=\hbar=1$ and denote the gravitational
constant $\kappa^2=8\pi G_N\equiv8\pi/M_{Pl}^2$, with the Planck mass being
$M_{PL}=G^{-1/2}_N=1.2\times 10^{19}\text{GeV}$.
We shall also write
\begin{equation}
F(R)=R+f(R)\,.\label{actiontwo}
\end{equation}
Thus, the modification to gravity is encoded in the function $F(R)$, which is added to the classical term $R$ of the Einstein-Hilbert action of General Relativity (GR). In what follows we will always discuss modified gravity in this form, by explicitly separating the contribution of GR from its modification.

Taking the variation of  Eq.~(\ref{action}) with respect to $g_{\mu\nu}$, one has:
\begin{equation}
R_{\mu\nu}-\frac{1}{2}Rg_{\mu\nu}=\frac{\kappa^2}{F'(R)}T^{{\mathrm{matter}}}_{\mu\nu}+\frac{1}{\kappa^2
F'(R)}\left\{\frac{1}{2}g_{\mu\nu}[F(R)-RF'(R)]
+(\nabla_{\mu}\nabla_{\nu}-g_{\mu\nu}\Box)F'(R)\right\}\,.\label{cucu}
\end{equation}
Here $R_{\mu\nu}$ is the Ricci tensor, the prime denoting the derivative with respect to the curvature scalar $R$, $\nabla_{\mu}$ is the covariant derivative operator associated with the metric $g_{\mu\nu}$, and $\Box\phi\equiv g^{\mu\nu}\nabla_{\mu}\nabla_{\nu}\phi$ is the
d'Alembertian of the scalar field $\phi$. Finally, $T^{{\mathrm{matter}}}_{\mu\nu}$ is the matter stress-energy tensor.
We will work in the spatially-flat Friedman-Robertson-Walker (FRW) space-time
described by the metric
\begin{equation}
ds^2=-dt^2+a^{2}(t)d {\bf x}^2\,,\label{FRWmetric}
\end{equation}
where $a(t)$ is the universe scale factor. The Ricci scalar is given by
\begin{equation}
R=6 \left(2H^2+\dot{H}\right)\,.\label{R}
\end{equation}

From (\ref{cucu}) the following gravitational field equations are obtained:
\begin{equation}
3 F'(R)H^2=\kappa^2\rho_{m}+\frac{1}{2}(F'(R)R-F(R))-3H\dot{F'}(R)\,, \label{uno}
\end{equation}
\begin{equation}
 -2F'(R)\dot{H}=\kappa^2(\rho_m+P_m)+\ddot{F}(R)-H\dot{F}(R)\,.\label{due}
\end{equation}
Here, $H=\dot{a}(t)/a(t)$ is the Hubble parameter and the dot denotes time
derivative $\partial/\partial t$, while $\rho$ and $p$ are, respectively, the matter
energy-density and pressure. The following matter conservation law results
\begin{equation}
\dot \rho_m+3H(\rho_m+p)=0\,.
\end{equation}
For a perfect fluid, it gives
\begin{equation}
p=\omega\rho\,,
\end{equation}
$\omega$ being the thermodynamical EoS-parameter for matter. For standard matter, $\omega=0$ and $\rho=a(t)^{-3}$ while, for radiation, $\omega=1/3$ and $\rho=a(t)^{-4}$.

Eqs.~(\ref{uno})-(\ref{due}) can be written as functions of the effective energy density, $\rho_{eff}$, and effective pressure, $p_{eff}$, of the universe. This yields
\begin{equation}
\rho_{\mathrm{eff}}=\frac{3}{\kappa^2}H^2\,,\label{FRW1}
\end{equation}
\begin{equation}
p_{\mathrm{eff}}=-\frac{1}{\kappa^2}\left(2\dot{H}+3H^2\right)\,,\label{FRW2}
\end{equation}
where
\begin{equation}
\rho_{\mathrm{eff}}=\frac{1}{F'(R)}\left\{\rho+\frac{1}{2\kappa^2}
\left[(F'(R)R-F(R))-6H\dot{F}'(R)\right]\right\}\,,\label{rho}
\end{equation}
\begin{equation}
p_{\mathrm{eff}}=\frac{1}{F'(R)}\left\{p+\frac{1}{2\kappa^2}\left[-(F'(R)R-F
(R))+4H\dot{F}'(R)
+2\ddot{F}'(R)\right]\right\}\,.\label{p}
\end{equation}
We note that Eqs.~(\ref{FRW1})-(\ref{FRW2}) have in fact the same form as the
Friedmann equations of General Relativity. Notice that the  part corresponding
to modified gravity has been formally included into $\rho_{eff}$ and $p_{eff}$.
Hence, one must not forget that gravitational terms enter here in both sides of the
FRW equation.

We will be interested in the cosmological behavior of realistic models of modified gravity describing the de Sitter epoch of the universe today. The tag `realistic' has been defined in previous papers and has to do with the feasibility of the models in view the all the most recent and accurate observational data. We define the dark energy density $\rho_{DE}$ as $\rho_{DE}=\rho_{eff}-\rho/F'(R)$ and introduce the variable\cite{Bamba}
\begin{equation}
y_H (z)\equiv\frac{\rho_{\mathrm{DE}}}{\rho_m^{(0)}}=\frac{H^2}{\tilde{m}^2}-(z+1)^3-\chi
(z+1)^{4}\,.\label{y}
\end{equation}
Here, $\rho_m^{(0)}$ is the energy density of matter at present time,
$\tilde{m}^2$ is the mass scale
\begin{equation}
\tilde{m}^2\equiv\frac{\kappa^2\rho_m^{(0)}}{3}\simeq 1.5 \times
10^{-67}\text{eV}^2\,,
\end{equation}
and $\chi$ is defined as
\begin{equation}
\chi\equiv\frac{\rho_r^{(0)}}{\rho_m^{(0)}}\simeq 3.1 \times
10^{-4}\,,
\end{equation}
where $\rho_r^{(0)}$ is the energy density of radiation at present, $z$ the redshift parameter, $z=1/a(t)-1$, and $y_H(z)$ is written as a function of $z$.

The EoS-parameter for dark energy, $\omega_{\mathrm{DE}}$, is
\begin{equation}
\omega_{\mathrm{DE}}=-1+\frac{1}{3}(z+1)\frac{1}{y_H(z)}\frac{d y_H(z)}{d (z)}\,.\label{oo}
\end{equation}
By combining Eq.~(\ref{FRW1}) with Eq.~(\ref{R}) and using
Eq.~(\ref{y}), one gets
\begin{equation}
\frac{d^2 y_H(z)}{d z^2}+J_1\frac{d y_H(z)}{d z}+J_2
y_H(z)+J_3=0\,,\label{superEq}
\end{equation}
where
\begin{equation}
J_1=\frac{1}{(z+1)}\left(-3-\frac{1}{y_H+(z+1)^{3}+\chi (z+1)^{4}}\frac{1-F'(R)}{6\tilde{m}^2
F''(R)}\right)\,,
\end{equation}
\begin{equation}
J_2=\frac{1}{(z+1)^2}\left(\frac{1}{y_H+(z+1)^{3}+\chi (z+1)^{4}}\frac{2-F'(R)}{3\tilde{m}^2
F''(R)}\right)\,,
\end{equation}
\begin{equation}
J_3=-3 (z+1)-\frac{(1-F'(R))((z+1)^{3}+2\chi (z+1)^{4})
+(R-F(R))/(3\tilde{m}^2)}{(z+1)^2(y_H+(z+1)^{3}+\chi
(z+1)^{4})}\frac{1}{6\tilde{m}^2
F''(R)}\,.
\end{equation}
Thus, we have
\begin{equation}
R=3\tilde{m}^2 \left(4y_H(z)-(z+1)\frac{d y_H(z)}{d z}+(z+1)^{3}\right)\,,\label{Ricciscalar}
\end{equation}
where we have used $-(z+1)H(z) d/d z=H(t) d/d(\ln a(t))=d/d t$.

In general, Eq.~(\ref{superEq}) can be solved in a numerical way, once we write the explicit form of the $F(R)$-model. Neglecting the contribution of matter, the trace of Eq.~(\ref{cucu}),
\begin{equation}
3\Box F'(R)+R F'(R)-2F(R)=0\,, \label{trace}
\end{equation}
 gives us the well-known de Sitter condition
\begin{equation}
2F(R_{dS})-R_{dS}F'(R_{dS})=0\,, \label{dS}
\end{equation}
where the Ricci scalar $R_{dS}$ is a constant.

As a check, let us now study perturbations around the de Sitter solution of the dark energy density, to see that we are able to recover well known results. Here we restrict our analysis to homogeneous perturbations. The behavior of general, linear, inhomogeneous perturbations has been discussed in \cite{fara}, where the equivalence between the two approaches has been
shown explicitly (see also the independent proof contained in \cite{cogno}).

The starting point will be
\begin{equation}
y_{H}(z)\simeq y_{0}+y(z)\,,\label{bum}
\end{equation}
where $y_0=R_{dS}/12\tilde{m}^2$ is a constant and $|y(z)|<<1$. Eq.~(\ref{Ricciscalar}) leads to
\begin{equation}
R=3\tilde{m}^2 \left(4y_0+4y(z)-(z+1)\frac{d y_H(z)}{d z}+(z+1)^3\right)\,.\label{RdS}
\end{equation}
In this case, by neglecting the contribution of radiation and assuming the matter one to be much smaller than $y_0$, Eq.~(\ref{superEq}) becomes
\begin{equation}
\frac{d^2 y(z)}{d z^2}+\frac{\alpha}{(z+1)}\frac{d y(z)}{d z}+\frac{\beta}{(z+1)^2}
y(z)=4\delta(z+1)\,,\label{superEqbis}
\end{equation}
where
\begin{equation}
\alpha=-2\,,\label{alpha}
\end{equation}
\begin{equation}
 \beta= -4+\frac{4F'(R_{dS})}{RF''(R_{dS})}\,,\label{beta}
\end{equation}
\begin{equation}
 \delta=1+\frac{1-F'(R_{dS})}{R_{dS}F''(R_{dS})}\,.\label{delta}
\end{equation}

Here we have performed the variation of Eq.~(\ref{RdS}) with respect to $R$, and we have used Eq.~(\ref{dS}) too. The solution of Eq.~(\ref{superEqbis}) is
\begin{equation}
y(z)=C_0(z+1)^{\frac{1}{2}\left(1-\alpha\pm\sqrt{(1-\alpha)^2-4\beta}\right)}+
\frac{4\delta}{\beta}(z+1)^3\,,\label{result}
\end{equation}
where $C_0$ is a constant. It is easy to see that $|y(z)|<<1$ when $z\rightarrow -1^{+}$, and, therefore, the de Sitter solution is stable provided that (see for instance {\cite{fara,barrow,monica})
\begin{equation}
\frac{F'(R_{dS})}{R_{dS}F''(R_{dS})}>1\,.\label{stability}
\end{equation}

We have two possible behaviors for the dark energy density in viable models of modified gravity, for a  stable de Sitter universe. If
\begin{equation}
\frac{25}{16}>\frac{F'(R_{dS})}{R_{dS}F''(R_{dS})}>1\,,\label{gamma>0}
\end{equation}
the solution approaches the de Sitter point as a power function of $(z+1)$, that is $y(z)\sim (z+1)^{\gamma}$, $\gamma>0$. Otherwise, if
\begin{equation}
\frac{F'(R_{dS})}{R_{dS}F''(R_{dS})}>\frac{25}{16}\,,\label{discriminant}
\end{equation}
the discriminant in the square root of Eq.~(\ref{result}) is negative and the
dark energy density shows an oscillatory behavior whose amplitude decreases as $(z+1)^{3/2}$, when $z\rightarrow -1^{+}$.
As a consequence, we can write $y_H(z)$ as
\begin{eqnarray}
\label{oscillatorysolution}&& y_H(z)= \frac{R_{dS}}{12\tilde{m}^2}+\left(\frac{1}{F'(R_{dS})-R_{dS}F''(R_{dS})}-1\right)(z+1)^3+
(z+1)^{\frac{3}{2}}\times\\ \nonumber
&& \left[A\cos\left(\sqrt{\left(\frac{4F'(R_{dS})}{R_{dS}F''(R_{dS})}-
\frac{25}{4}\right)}\log(z+1)\right)+B\sin\left(\sqrt{\left(\frac{4F'(R_{dS})}{R_{dS}F''(R_{dS})}-
\frac{25}{4}\right)}\log(z+1)\right)\right]\,,
\end{eqnarray}
$A$ and $B$ being constants which depend on the boundary conditions.

Using Eq.~(\ref{oo}), we can evaluate the $\omega_{DE}$ parameter
\begin{equation}
 \omega_{DE}=-1+\frac{4\delta}{\beta}\frac{(z+1)^3}{y_0}+\frac{1}{3}\gamma\frac{(z+1)^\gamma}{y_0}\,,
\end{equation}
where
\begin{equation}
 \gamma=\frac{1}{2}\left(1-\alpha\pm\sqrt{(1-\alpha)^2-4\beta}\right)\,.\label{gamma}
\end{equation}
In the case of oscillating models which satisfy Eq.~(\ref{discriminant}), one has
\begin{eqnarray}
\label{omegaoscillating}&&\omega_{DE}=-1+\frac{12\tilde{m}^2}{R_{dS}}\left(\frac{1}{F'(R_{dS})-
R_{dS}F''(R_{dS})}-1\right)(z+1)^3+4\tilde{m}^2\frac{(z+1)^{\frac{3}{2}}}{R_{dS}}\times\\\nonumber
&&\left[A'\cos\left(\sqrt{\left(\frac{4F'(R_{dS})}{R_{dS}F''(R_{dS})}-\frac{25}{4}\right)}
\log(z+1)\right)+B'\sin\left(\sqrt{\left(\frac{4F'(R_{dS})}{R_{dS}F''(R_{dS})}-
\frac{25}{4}\right)}\log(z+1)\right)\right]\,,
\end{eqnarray}
with
\begin{equation}
A'=\frac{3}{2}A+\sqrt{\left(\frac{4F'(R_{dS})}{R_{dS}F''(R_{dS})}-\frac{25}{4}\right)}B\,,\label{o}
\end{equation}
and
\begin{equation}
B'=\frac{3}{2}B-\sqrt{\left(\frac{4F'(R_{dS})}{R_{dS}F''(R_{dS})}-\frac{25}{4}\right)}A\,. \label{u}
\end{equation}
We observe that $\omega_{DE}$ exhibits the same oscillation period of $y_H(z)$ and that its amplitude is amplified by its frequency, written in the coefficients $A'$ and $B'$.

In the case of models of Eq.~(\ref{actiontwo}), the oscillatory condition leads to
\begin{equation}
 \frac{1+f'(R_{dS})}{R_{dS} f''(R_{dS})}>\frac{25}{16}\,.\label{OC}
\end{equation}
The large class of one-step models \cite{Sawiki, onestep} which mimic the cosmological constant in the high-curvature regime (that is, when $f(R_{dS})\simeq-2\Lambda$, where $\Lambda$ is the cosmological constant) satisfy this condition ($f''(R_{dS})<<1$) and the corresponding $\omega_{DE}$ oscillates around the value $-1$.

\subsection{Time evolution}

% We are interested in the so-called IV Types finite-time singularities. A finite-time singularity occurs when Hubble parameter or some its derivatives (and therefore energy density, scale factor or, more simplicity, the Riemann tensor components) diverge in a finite time into the future. The softer kinds of such singularities are the Type IV, giving by the Hubble parameter
% \begin{equation}
% H=H_0+h_0(t_0-t)^\beta\,,
% \end{equation}
% where $\beta$ is a constant so that $\beta>1$. Here, $h_0$, $H_{0}$ and $t_{0}$ are positive constants and the cosmic time $t$ has to be $t<t_{0}$ because it should be for expanding universe. When $t$ is close to $t_{0}$, some high-order derivatives of $H$ and therefore the curvature become singular. We note that in the case of IV Type singularities the Hubble parameter and the Ricci scalar $R$ (which depends on $H$ and $\dot{H}$) do not diverge.

Let us now consider the stable de Sitter solution of Eqs.~(\ref{bum}) and (\ref{result}) in the case of $\gamma$ being a positive real number (what means that the condition (\ref{gamma>0}) is satisfied):
\begin{equation}
 y_H(z)=y_0+ C_0(z+1)^\gamma+\frac{4\delta}{\beta}(z+1)^3\,.
\end{equation}
Here $\beta$, $\delta$ and $\gamma$ have been given by Eqs.~(\ref{beta})-(\ref{delta}) and (\ref{gamma}),
$y_0=H_{dS}^2/\tilde{m}^2=R_{dS}/12\tilde{m}^2$ and $C_0$ is a constant. We will assume $C_0>0$.
The first equation of motion (\ref{FRW1}) leads to
\begin{equation}
 \frac{H^2}{\tilde{m}^2}=y_H(z)+(z+1)^3\equiv y_0+ C_0(z+1)^\gamma + \left(\frac{1}{F'(R_{dS})-R_{dS}F''(R_{dS})}\right)(z+1)^3\,.
\end{equation}
We will explicitly solve $H$ as a function of the cosmic time $t$. By writing $z+1$ as $1/a(t)$, one gets
\begin{equation}
\left(\frac{\dot{a}(t)}{a(t)}\right)^2=H_{dS}^2+(C_0\tilde{m}^2)\left(\frac{1}{a(t)}\right)^\gamma\,.
\end{equation}
We have here omitted the matter contribution to $y_H(z)$. By considering $t>0$, the general solution for the expanding universe is
\begin{equation}
 a(t)=\left(\frac{C_0\tilde{m}^2}{H^2_{dS}}\right)^\frac{1}{\gamma}\left[\sinh\left(\frac{H_{dS}}{2}\gamma t+\phi\right)\right]^{\frac{2}{\gamma}}\,,
\end{equation}
being $\phi$ a positive constant. It is then easy to obtain
\begin{equation}
 a(t)=a_0e^{H_{dS}t}\left[1-e^{-\left(\frac{H_{dS}}{2}\gamma t+\phi\right)}\right]^\frac{2}{\gamma}\,,
\end{equation}
where $a_{0}$ is a constant which depends on $\phi$. As $\gamma>0$, we get $a(t)\simeq a_0e^{H_{dS} t}$. For the Hubble parameter, one has
\begin{equation}
H=H_{dS}\coth\left(\frac{1}{2}H_{dS}\gamma t+\phi\right)\,,
\end{equation}
and, in general, for $t>0$ it is $H\simeq H_{dS}$.

Consider now the case of an oscillatory behavior followed by the condition (\ref{discriminant}). Then
\begin{equation}
H=\left(\frac{\dot{a}(t)}{a(t)}\right)=\sqrt{H_{dS}^2+
\tilde{m}^2a(t)^{-\frac{3}{2}}\left(A\cos\left(\mathcal{F}\log(a(t)^{-1})\right)+
B\sin\left(\mathcal{F}\log(a(t)^{-1})\right)\right)}\,,\label{bip}
\end{equation}
where we have used Eq.~(\ref{oscillatorysolution}), omitting matter contributions, and the frequency $\mathcal{F}$ is
\begin{equation}
\mathcal{F}=\sqrt{\frac{4F'(R_{dS})}{R_{dS}F''(R_{dS})}-\frac{25}{4}}\,.
\end{equation}
Assuming $a(t)\simeq \exp(H_0 t)$, Eq.~(\ref{bip}) yields
\begin{equation}
H\simeq\sqrt{H_{dS}^2+\tilde{m}^2 \left(\frac{1}{e^{H_0 t}}\right)^{\frac{3}{2}}\left[-A\cos\left(\mathcal{F} H_{0}t\right)+B\sin\left(\mathcal{F} H_0 t\right)\right]}\,.
\end{equation}
Also, in this case, $H\simeq H_{dS}$.
We should stress that, as
\begin{equation}
\frac{d^n}{dt^n}H(t)=\left(-H(z)(z+1)\frac{d}{dz}\right)^n H(z)\,,
\end{equation}
it is quite simple to see that any derivative of $H(z)$ becomes singular around the zeros of the sinus and cosinus functions.

\subsection{Singularities and final attractors}

We know, that if condition (\ref{stability}) is satisfied, the de Sitter point is stable, and it may be a final attractor of the system. However, in many papers, it has been demonstrated that the presence of finite-time singular solutions in expanding universe influences this behaviour. Finite-time singularities appear when Hubble parameter has the form
\begin{equation}
H=\frac{h_0}{(t_{0}-t)^{\beta}}\,,
\label{Hsingular}
\end{equation}
where $h_0$ is a positive constants and $t<t_{0}$.
If
$\beta>-1$, $H$ or its derivative and therefore the Ricci scalar diverge and become singular at the time $t_0$.

If a modified gravity theory shows such kind of singularity where $R\rightarrow\pm\infty$, the cosmological expansion could tend towards such asymptotic solution and the universe becomes singular. This is because singular solutions often are energetically accessible for the system and destabilize the model. As a qualitative example, we can consider the case of a realistic $f(R)$-model, namely the Hu-Sawiki Model~\cite{Sawiki}, able to reproduce the de Sitter phase of our universe,
\begin{equation}
F(R)=R-\frac{\tilde{m}^{2}c_{1}(R/\tilde{m}^{2})^{n}}{c_{2}(R/\tilde{m}^{2})^{n}+1}\equiv R-\frac{\tilde{m}^{2}c_{1}}{c_{2}}+\frac{\tilde{m}^{2}c_{1}/c_{2}}{c_{2}(R/\tilde{m}^{2})^{n}+1}\,.\label{HuSawModel}
\end{equation}
Here, $\tilde{m}^{2}$ is a mass scale, $c_{1}$ and $c_{2}$ are positive parameters and $n$ is a natural positive number.
The model is very carefully constructed, such that $c_{1}\tilde{m}^{2}/c_{2}\simeq2\Lambda$, where $\Lambda$ is the Cosmological Constant, and in the high curvature region the physic of $\Lambda$CDM Model can be found.
The Hu-Sawiki Model could become singular when $R$ diverges and $\beta=-n/(n+2)>-1$.

Let us rewrite Eq.(\ref{trace}) neglecting the contribute of matter in the well known form
\begin{equation}
\Box F'(R)=\frac{\partial V_{\mathrm{eff}}}{\partial F'(R)}\,,\label{scalaroneeqbis}
\end{equation}
where
\begin{equation}
\frac{\partial V_{\mathrm{eff}}}{\partial
F'(R)}=\frac{1}{3}\left[2F(R)-RF'(R)\right]\,.\label{Veff}
\end{equation}
Here, $F'(R)$ is the so-called `scalaron', this is the effective scalar degree
of freedom of modified gravity, and $V_{\mathrm{eff}}$ is the effective potential associated with the scalaron.
In the case of Hu-Sawiki Model, the scalaron reads
\begin{equation}
F'(R)=1-\frac{\tilde{m}^{2}c_{1}/c_{2}}{\left(c_{2}(R/\tilde{m}^{2})^{n}+1\right)^2}(n)\frac{c_{2}R^{n-1}}{(\tilde{m}^2)^n}\,.
\end{equation}
It tends to a constant when $R\rightarrow\pm\infty$. Furthermore, by writing $\partial V_{\rm eff}/\partial R$ as $F''(R)(\partial V_{\rm eff}/\partial F'(R))$, in principle one can evaluate the potential $V_{\rm eff}$ of Eq.~(\ref{scalaroneeqbis}) through an integration. When $R\rightarrow\pm\infty$ one easily finds
\begin{equation}
V_{\rm eff}(R\rightarrow\pm\infty)\simeq-\frac{\tilde{m}^{2}c_{1}/c_{2}}{3c_2(R/\tilde{m}^2)^n}(n+1)\,.
\end{equation}
The Hu-Sawiki Model exhibits a stable de Sitter solution in vacuum, that may be the final attractor of the system. However,
we can observe that if a singular solution with $R$ diverging exists, it is at a finite value of $V_{\rm eff}$ (in particular, it tends to zero) and the scalaron $F'(R)$ can crossover the potential in some point of cosmological evolution and arise the value $F'(R)=0$ for which catastrophic curvature singularity emerges. In general, it is possible to see that singularities violate the strong energy condition (SEC) describing acceleration. This is the reason for which realistic models of modified gravity describing the current acceleration of Universe could become unstable and fall into a singularity. The study of the singularities is fundamental in order to achieve a correct description of the Universe. As regard this argument, we will make a confront with realistic exponential model in the last section.

\section{$\omega_{DE}$-oscillations in the matter era}

Let us now consider matter era solutions. The trace of Eq.~(\ref{cucu}) reads, in this case,
\begin{equation}
\Box F'(R)=\frac{1}{3}\left(2F(R)-RF'(R)-\kappa^2\rho_m\right)\,,
\end{equation}
where we have explicitly written the standard matter contribution to the stress energy tensor $T^{\mathrm{matter}}_{\mu\nu}$.

The critical points associated with matter dominated era for  a generic $F(R)$ model have been investigated in Ref.
\cite{Amendola}. The results are
\begin{eqnarray}
\dot{F'}(R)&=&0\,, \quad \rho_r=0\,,  \quad R=3H^2\, \\
F(R)&=&RF'(R)\,.\label{mattercondition}
\end{eqnarray}
As a consequence, from  Eq.~(\ref{rho}) around the critical points, one has  $\rho_{eff}\simeq\rho_{matter}/F'(R)$.
%around the critical point
%$R\simeq 3\tilde{m}^2(z+1)^3$ and
%\begin{equation}
%|R F'(R)-F(R)|\ll 6\tilde{m}^2(z+1)^3\,.
%\end{equation}

Now we  consider the perturbations of dark energy during the matter epoch, by assuming that $y_H(z)\ll(1+z)^3$. Eq.~(\ref{superEq}) reads, to first order in $y_H(z)/(z+1)^3$,
\begin{eqnarray}
\label{Trento}&& y''_H(z)-\frac{y'_H(z)}{(z+1)}\left(3\right)+
\frac{y(z)}{(z+1)^2}\left(\frac{4F'(R)-3}{R F''(R)}\right)=\\\nonumber
&& (z+1)\left[3+\frac{1}{2F''(R)R}\left((1-F'(R))+\frac{(R-F(R))}{R}\right)\right]\,,
\end{eqnarray}
where $R$ is written in full form, as in Eq.~(\ref{Ricciscalar}), and we have used the conditions (\ref{mattercondition}). Here and in what follows we have put $R-F(R)\simeq 6\tilde{m}^2y_H(z)$, as a consequence of Eq.~(\ref{rho}) and conditions (\ref{mattercondition}).

In the standard cosmological scenario, the effects of dark energy are completely neglected when $z\gg 3$. In this case the Ricci scalar simply reads
\begin{equation}
 R=3\tilde{m}^2(z+1)^3\,,
\end{equation}
and Eq.~(\ref{Trento}) can be expanded as
\begin{eqnarray}
\label{strong}&&y_H''(z)+y'_H(z)\frac{1}{(z+1)}\left(-\frac{7}{2}-\frac{(1-F'(R))F'''(R)}{2F''(R)^2}\right)+
\\\nonumber &&y_H(z)\frac{1}{(z+1)^2}\left(2+\frac{1}{R F''(R)}+\frac{2(1-F'(R))F'''(R)}{F''(R)^2}\right)=\left(3+\frac{2-F'(R)-F(R)/R}{2RF''(R)}\right)(z+1)\,.
\end{eqnarray}
In order to solve Eq.~(\ref{strong}) we can set $z=z_0+(z-z_0)$, where $|z-z_0|\ll z_0$, and perform a variation with respect to $z$. To first order in $(z-z_0)$, we find
\begin{eqnarray}
&&y_H''(z)+y'_H(z)\frac{1}{(z_0+1)}\left(-\frac{7}{2}-\frac{(1-F'(R_0))F'''(R_0)}{2F''(R_0)^2}\right)+
\\\nonumber &&y_H(z)\frac{1}{(z_0+1)^2}\left(2+\frac{1}{R_0 F''(R_0)}+\frac{2(1-F'(R_0))F'''(R_0)}{F''(R_0)^2}\right)=\\\nonumber
&&\left(3+\frac{2-F'(R_0)-F(R_0)/R_0}{2R_0F''(R_0)}\right)(z_0+1)+\\\nonumber
&& 3\left(\frac{1}{2}+\frac{5F(R_0)/R_0-F'(R_0)-4}{6R_0F''(R_0)}-
\frac{(2-F'(R_0)-F(R_0)/R_0)F'''(R_0)}{2F''(R_0)^2}\right)(z-z_0)\,,
\end{eqnarray}
where $R_0=3\tilde{m}^2(z_0+1)^3$.
%In this formula, we are considering $y_H(z)$ as $y_H(z)=y(z_0)+y(z-z_0)$ with $y_H(z_0)=const$ (local approximation).
The solution of this equation is
\begin{equation}
y_H(z)= a+b\cdot(z-z_0) + C\cdot e^{\frac{1}{2(z_0+1)}\left(\alpha\pm\sqrt{\alpha^2-4\beta}\right)(z-z_0)}\,, \label{resultmatter}
\end{equation}
where $C$ is constant and
\begin{eqnarray}
&& a=\left(\frac{1}{6\tilde{m}^2}\right)\frac{6R_0^2F''(R_0)+(2-F'(R_0))R_0-F(R_0)}{1+2R_0F''(R_0)
+2(2-F'(R_0)-F(R_0)/R_0)R_0 F'''(R_0)/F''(R_0)}+\\\nonumber
&&\left(\frac{R_0^2}{4\tilde{m}^2}\right)\frac{7F''(R_0)^2+(2-F'(R_0)-F(R_0)/R_0)
F'''(R_0)}{[2R_0F''(R_0)^2+F''(R_0)+2R_0(2-F'(R_0)-F(R_0)/R_0)F'''(R_0)]^2}\times\\\nonumber
&&[RF''(R_0)^2+(5F(R_0)/R_0-F'(R_0)-4)F''(R_0)/3- R_0(2-F'(R_0)-F(R_0)/R_0)F'''(R_0)]\,,  \label{a}
\end{eqnarray}
\begin{equation}
b=\frac{R_0}{2\tilde{m}^2(z_0+1)}\, \frac{R_0F''(R_0)^2+(5F(R_0)/R_0-F'(R_0)-4)F''(R_0)/3
-(2-F'(R_0)-F(R_0))R_0 F'''(R_0))}{2R_0F''(R_0)^2+F''(R_0)+2(2-F'(R_0)-F(R_0)/R_0)R_0 F'''(R_0)}\,, \label{b}
\end{equation}
\begin{equation}
\alpha=\frac{7}{2}+\frac{(1-F'(R_0))F'''(R_0)}{2F''(R_0)^2}\,,\label{1}
\end{equation}
\begin{equation}
\beta=2+\frac{1}{R_0 F''(R_0)}+\frac{2(1-F'(R_0))F'''(R_0)}{F''(R_0)^2}\,.\label{2}
\end{equation}
Let us now analyze this result. Since in the expanding universe $(z-z_0)<0$, it turns out that the matter solution is stable around $R_0$ if $\alpha>0$ and $\beta>0$. This means that
\begin{equation}
\frac{(1-F'(R))F'''(R_0)}{2F''(R_0)^2}>-\frac{7}{2}\,,\label{qq}
\end{equation}
\begin{equation}
 \frac{1}{R_0F''(R_0)}>12\,.\label{q}
\end{equation}
We can thus have an oscillatory behavior of the dark energy if the discriminant of the square root of Eq.~(\ref{resultmatter}) is negative.

\subsection{Late-time matter era}

The effects of dark energy could be relevant at a late-time matter era, near the transition between the matter and de Sitter epochs ($1\lesssim z\lesssim 3$). In this case, we cannot do an expansion of the $F(R)$-functions in terms of $y_H(z)$, as we did before. On the other hand, in realistic models of modified gravity, $y_H(z)$ tends to a constant value, as in Eq.~(\ref{bum}), $y_H(z)=y_0+y(z)$, where $y_0\simeq R_{dS}/12\tilde{m}^2$ is related to the de Sitter solution and $|y(z)|\ll y_0$ (in this way, we can reproduce the correct dynamical evolution of the universe, as in the $\Lambda$CDM model). As a consequence, we can actually perform the variation of Eq.~(\ref{Trento}) with respect to $y(z)$, to obtain
\begin{eqnarray}
&&y''(z)+y'(z)\frac{1}{(z+1)}\left[-\frac{7}{2}-\frac{(1-F'(R))F'''(R)}{2F''(R)^2}\right]+
\frac{y_0+y(z)}{(z+1)^2}\, \frac{4F'(R)-3}{RF''(R)}=\\\nonumber
&& (z+1)\left[3+\frac{1}{2F''(R)R}\left((1-F'(R))+\frac{(R-F(R))}{R}\right)\right]\,,
\end{eqnarray}
where
\begin{equation}
R=3\tilde{m}^2\left[(z+1)^3+4y_0\right]\,.
\end{equation}
Also in this case, we can take $z=z_0+(z-z_0)$, where $|z-z_0|\ll z_0$, and doing the variation with respect to $z$, we find, up to first order in $(z-z_0)$,
\begin{eqnarray}
&&y''(z)+y'(z)\frac{1}{(z_0+1)}\left[-\frac{7}{2}-\frac{(1-F'(R_0))F'''(R_0)}{2F''(R_0)^2}\right]+
\frac{y_0+y(z)}{(z_0+1)^2}\, \frac{1}{R_0 F''(R_0)}=\\\nonumber
&& (z_0+1)\left[3+\frac{1}{2F''(R_0)R_0}\left(1-F'(R_0)+\frac{R_0-F(R_0)}{R_0}\right)\right]+\\ \nonumber
&& 3\left[\frac{1}{2}-\frac{1-F'(R_0)}{2F''(R_0)^2}F'''(R_0)+
\frac{1-F'(R_0)}{6F''(R_0)R_0}\right](z-z_0)\,,
% &&
% \left(\frac{F'''(R_0)}{F''(R_0)^2}+\frac{1}{R_0F''(R_0)}\right)(1-F(R_0)/R_0)^2)(z-z_0)\,,
\end{eqnarray}
where $R_0=3\tilde{m}^2((z_0+1)^3+4y_0)$. In this expression we have used the condition (\ref{mattercondition}) and we have set $R_0-F(R_0)\simeq 6\tilde{m}^2y_0$. Owing to the fact that $y_0\ll R_0$, the variations with respect to $y_0$ are at least of first order in $y_0/R_0$. The solution of this equation is
\begin{eqnarray}
&&y_0=a\,,\nonumber\\
&&y(z)= b\cdot(z-z_0) + C\cdot e^{\frac{1}{2(z_0+1)}\left(\alpha\pm\sqrt{\alpha^2-4\beta}\right)(z-z_0)}\,, \label{resultmatter2}
\end{eqnarray}
where $C$ is a constant, and
\begin{eqnarray}
&&\label{eins}a\simeq\frac{R_0}{6\tilde{m}^2(4F'(R_0)-3)}(6F''(R_0)R_0+2-F'(R_0)-F(R_0)/R_0)+\\ \nonumber
&&\frac{R_0^2}{4\tilde{m}^2}(7F''(R_0)^2+2-F'(R_0)-F(R_0)/R_0)F'''(R_0))\times\\ \nonumber
&&(R_0F''(R_0)^2-(1-F'(R_0))R_0F'''(R_0)+(1-F'(R_0))F''(R_0)/3)\,,\\
&&\label{zwei}b=\frac{3(z_0+1)^2(R_0F''(R_0)^2-(1-F'(R_0))R_0F'''(R_0)+(1-F'(R_0))F''(R_0)/3)}{2\, F''(R_0)}\nonumber\\
%&&+\frac{3(z_0+1)^2(R_0F'''(R_0)+F''(R_0))(1-F(R_0)/R_0)^2}{F''(R_0)}\,,\\
&&\label{drei}\alpha=\frac{7}{2}+\frac{(1-F'(R_0))F'''(R_0)}{2F''(R_0)^2}\,,\\
&&\label{vier}\beta=\frac{1}{R_0 F''(R_0)}\,.
\end{eqnarray}
%In computation of $a$, we avoid the corrections of $y_0/R_0$.
The solution is stable around $R_0$ if $\alpha>0$ and $\beta>0$. This means that
\begin{equation}
\frac{(1-F'(R_0))F'''(R_0)}{2F''(R_0)^2}>-\frac{7}{2}\,,\label{x}
\end{equation}
\begin{equation}
 \frac{1}{R_0F''(R_0)}>0\,.\label{xx}
\end{equation}
The oscillatory behavior of the dark energy occurs when the discriminant of the square root of Eq.~(\ref{resultmatter2}) is negative.

We observe that the expression (\ref{resultmatter2}) is more accurate than (\ref{resultmatter}). In general, if the conditions (\ref{x})-(\ref{xx}) are satisfied for $R_0=3\tilde{m}^2(z_0+1)^3+12\tilde{m}^2y_0$, the conditions (\ref{q})-(\ref{qq}) will be also satisfied, provided it is possible to use the approximation $R_0=3\tilde{m}^2(z_0+1)^3$.
In particular, since $F'(R)\simeq 1$, the condition
 \begin{equation}
  \frac{1}{F''(R)}>>1\,,
 \end{equation}
 is sufficient to obtain stability during the matter era and an oscillating behavior of the dark energy. Such condition has to be satisfied by models which mimic the $\Lambda$CDM Model ($F(R)\simeq R-2\Lambda$), for which $F''(R)\rightarrow 0^+$.

\subsection{$\omega_{DE}$-parameter in the matter era}

We can use the solutions (\ref{resultmatter}) or (\ref{resultmatter2}) and the expression (\ref{oo}) to evaluate the parameter $\omega_{DE}$ around $z=z_0$ during the matter era. We get
\begin{equation}
\omega_{DE}\simeq-1+ \frac{1}{3a}\left[b(z_0+1)+C\frac{\alpha\pm\sqrt{\alpha^2-4\beta}}{2}e^{\frac{1}{2(z_0+1)}
\left(\alpha\pm{\sqrt{\alpha^2-4\beta}}\right)(z-z_0)}\right]\,.
\end{equation}
In the case of the oscillating models, for which $\alpha^2-4\beta<0$, this equation reads
\begin{equation}
 \omega_{DE}\simeq -1+\frac{b}{3a}(z_0+1)+\frac{e^{\frac{\alpha}{2(z_0+1)}}}{6a}\left[A\cos \frac{\sqrt{4\beta-\alpha^2}}{2(z_0+1)}(z-z_0)+B\sin \frac{\sqrt{4\beta-\alpha^2}}{2(z_0+1)}(z-z_0)\right]\,, \label{omegaoscillatingdue}
\end{equation}
where $A$ and $B$ are constant, proportional to the period $\sqrt{4\beta-\alpha^2}/2(z_0+1)$.

\section{Oscillating behavior of the dark energy in a exponential model}

In Refs.~\cite{Sawiki,onestep,Battye} several versions of viable modified
gravity have been proposed, namely models, which are able to
reproduce the current acceleration of the universe within the bounds set by
the most recent observational data. They incorporate
a vanishing (or fast decreasing) cosmological constant in the flat
($R\rightarrow 0$) limit, and exhibit a suitable, constant asymptotic
behavior for large values of $R$. The simplest one was proposed in
Ref.~\cite{onestep} and can be given under the form of Eq.~(\ref{actiontwo}), namely
$
F(R)=R+f(R)\,
$,
where
\begin{equation}
f(R)=-2\Lambda(1-\mathrm{e}^{-R/R_0})\,.
\end{equation}
Here, $\Lambda\simeq 10^{-66}\text{eV}^2$ is the cosmological
constant and
$R_{0}\sim \Lambda$ a curvature parameter. In flat space, one has
$f(0)=0$
and recovers the Minkowski solution. For $R \gg R_{0}$, $f(R)\simeq
-2\Lambda$, and the theory mimics the $\Lambda$CDM model. The viability of
this kind of $\Lambda$ models has been studied in Ref.~\cite{onestep,twostep}.
We are allowed to set
\begin{equation}
 R_0= b\, \Lambda\,,
\end{equation}
 where $b>0$ needs to be sufficiently small, in order to find the de Sitter solution $R_{dS}$ at high curvature ($f(R_{dS})\simeq -2\Lambda$), so that de Sitter condition (\ref{dS}) leads
\begin{equation}
 R_{dS}\simeq 4\Lambda\,,
\end{equation}
in accordance with the $\Lambda$CDM model. The stability condition (\ref{stability}) yields
\begin{equation}
\frac{b^2 e^\frac{4}{b}}{2}-b>4\,.
\end{equation}
It is easy too see that this condition is always satisfied and, in particular, the oscillating condition of Eq.~(\ref{OC}) is satisfied for any value of $b$. In what follows, we will consider the case $b=1$.

Eq.~(\ref{superEq}) can be solved in a numerical way. To this purpose we have imposed the following initial conditions at $z=z_i$:
\begin{equation*}
\frac{d y_H}{d (z)}\Big\vert_{z_i}=0\,,
\end{equation*}
\begin{equation}
y_H\Big\vert_{z_i}=\frac{\Lambda}{3\tilde{m}^2}\,,
\end{equation}
which correspond to the ones of the $\Lambda$CDM model. This choice obeys to
the fact that in the high redshift regime the exponential model is quite
close to the $\Lambda$CDM Model. The value of $z_i$ has been chosen so
that $RF''(z = z_i) \sim 10^{-7}$, assuming $R= 3\tilde{m}^2 (z+1)^3+4\Lambda$, and $z_i=2.6$.
In setting the parameters we have used the last published results of the
$W$MAP, BAO, and SN surveys \cite{WMAP} and also that $\Lambda=(7.93)\tilde{m}^2$.

A plot of $y_H(z)$ is shown in Fig.~1. In order to study the behavior of dark energy, we also extrapolate the value of the corresponding density parameter $\Omega_\mathrm{DE}(z)$, as
\begin{equation}
\Omega_\mathrm{DE}(z)\equiv\frac{\rho_\mathrm{DE}}{\rho_\mathrm{eff}}
=\frac{y_H(z)}{y_H(z)+\left(z+1\right)^3+\chi\left(z+1\right)^4}\,.
\end{equation}
In Fig.~2, the corresponding plot of $\Omega_\mathrm{DE}(z)$ is overlapped with the one for $y_H(z)-\Lambda/3\tilde{m}^2$.

\begin{figure}[-!h]
\begin{center}
\includegraphics[angle=0, width=0.55\textwidth]{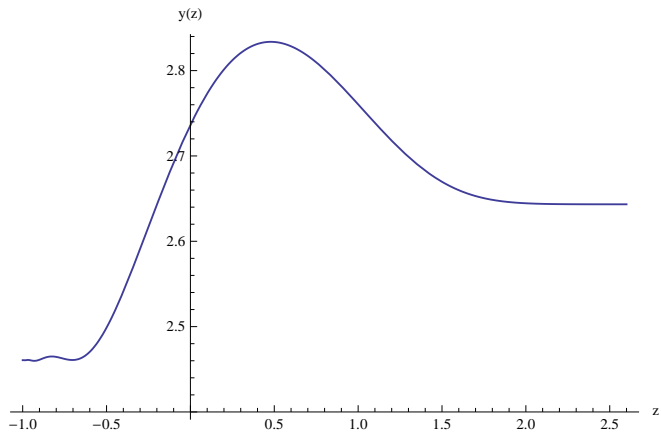}
\end{center}
%  \begin{center}
\caption{Plot of $y_H(z)$.\label{Fig1}}
%\end{center}
\end{figure}
\begin{figure}[-!h]
\begin{center}
\includegraphics[angle=0, width=0.55\textwidth]{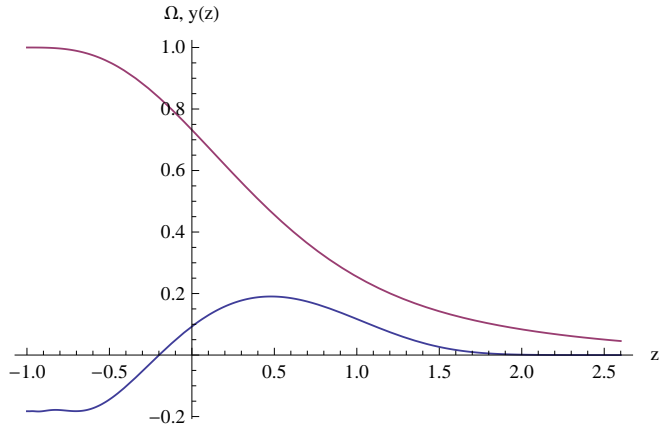}
\end{center}
%  \begin{center}
\caption{Plot of $\Omega_\mathrm{DE}(z)$ overlapped with $y_H(z)-\Lambda/3\tilde{m}^2$.\label{Fig2}}
%\end{center}
\end{figure}

We must point out that the data are in perfect accordance with the most recent, accurate
observations of our current universe, where
\begin{equation*}
\Omega_\mathrm{DE}(0)=0.721\pm 0.015\,.
\end{equation*}
In fact, this value is to be compared with what we get at present red shift, $z=0$, namely \, $\Omega_\mathrm{DE}(0)=0.732$.

\subsection{De Sitter oscillations}

Dark energy is dominant near the de Sitter solution describing the present acceleration of the universe  (for $z<1.5$, $\Omega_{DE}>0.5$). As the de Sitter solution is a stable point of the theory, it turns out that $y_{H}(z\rightarrow -1)\simeq R_{dS}/12\tilde{m}^2$, where $R_{dS}$ is the solution of Eq.~(\ref{dS}). In our case this equation is transcendental and such that taking $f(R_{dS})\simeq f(4\Lambda)$, one has
\begin{equation}
R_{dS}\simeq 4\Lambda\times(0.947)\,.
\end{equation}
A computational evaluation (extrapolated from $y_H(z=0.9)$) gives $R_{dS}\simeq 3.725\Lambda$ so that we are remarkably close to the corresponding value for the $\Lambda$CDM model, namely $R_{dS}=4\Lambda$. Further, as the condition (\ref{discriminant}) is satisfied, we can predict an oscillatory behavior of $y_H(z)$, which is shown in fact in Fig.~1. The value of $\omega_{DE}$ oscillates infinitely often around the line of phantom divide at $\omega_{DE}=-1$, as in Eq.~(\ref{omegaoscillating}). In Fig.~3 the behavior of $\omega_{DE}$ is depicted. In order to appreciate the constant frequency of $\omega_{DE}$ with respect to the logarithmic scale during the de Sitter phase, in Fig.~4 we plot the values of
\begin{equation}
\tilde{\omega_{DE}}(\log(z+1))=\left[\omega_{DE}+1-\frac{12\tilde{m}^2}{R_{dS}}
\left(\frac{1}{F'(R_{dS})-R_{dS}F''(R_{dS})}-1\right)(z+1)^3\right](z+1)^{-\frac{3}{2}}\,,
\end{equation}
as a function of $\log(z+1)$, for $-0.999<z<0$ (here, $R_{dS}=3.725\Lambda$). In this way, we stress the oscillating part of Eq.~(\ref{omegaoscillating}), whose frequency is proportional to $2\pi/\sqrt{4/(R_{dS}f''(R_{dS}))-25/4}\simeq 1.570$. As last point, we should remark that the amplitude of $\omega_{DE}$ is amplified with respect to $y_H(z)$ by its frequency, and it decreases as $(z+1)^{3/2}$. These data are also in good accordance with the observations of our present universe, which yield
\begin{equation}
\omega_{DE}=-0.972^{+0.061}_{-0.060}\,.
\end{equation}

\begin{figure}[-!h]
\begin{center}
\includegraphics[angle=0, width=0.55\textwidth]{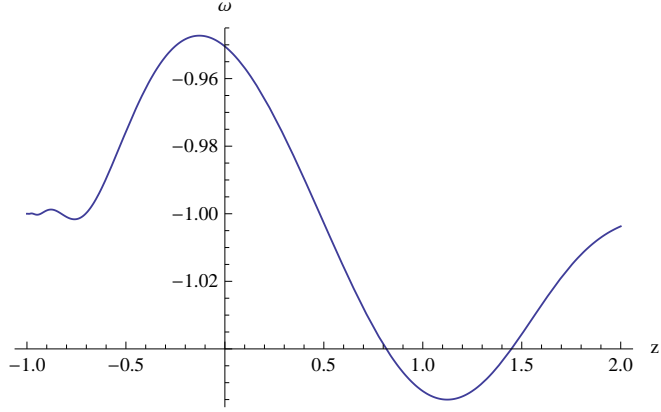}
\end{center}
%  \begin{center}
\caption{Plot of $\omega_{DE}$ for $-1<z<2$.\label{Fig3}}
%\end{center}
\end{figure}
\begin{figure}[-!h]
\begin{center}
\includegraphics[angle=0, width=0.55\textwidth]{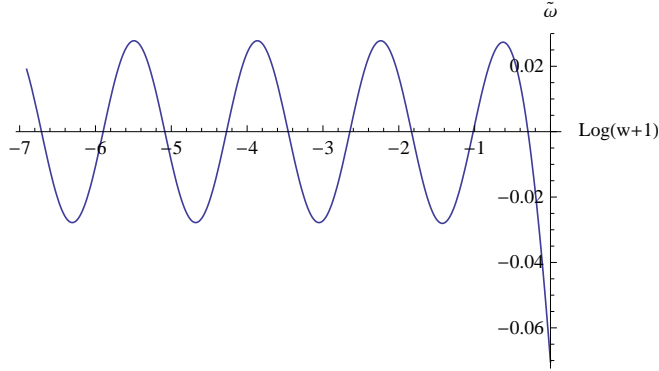}
\end{center}
%  \begin{center}
\caption{Plot of $\tilde{\omega}_{DE}$ for $\log[-0.999+1]<\log(z+1)<\log[0+1]$.\label{Fig4}}
%\end{center}
\end{figure}

\subsection{Matter oscillations}

In the high curvature region matter is dominant ($z>1.5$ with $\Omega_{DE}(z)\sim 0.1$) and $y_H\simeq \Lambda/3\tilde{m}^2$, as it is clear in Fig.~2. Eq.~(\ref{resultmatter2}) yields an estimation of the dark energy density value around $z_0$. Owing to the fact that, in this region, $f''(R)$ is very close to $0^+$, dark energy oscillates as
\begin{equation}
y_H(z)= a+e^{\frac{\alpha(z-z_0)}{2(z_0+1)}}\left[A\sin\left(\frac{\sqrt{\beta}}{(z_0+1)}(z-z_0)\right)+
B\cos\left(\frac{\sqrt{\beta}}{(z_0+1)}(z-z_0)\right)\right]\,,
\end{equation}
where $A$ and $B$ are constant, and
\begin{equation}
a\simeq \Lambda/3\tilde{m}^2\,,
\end{equation}
\begin{equation}
\alpha=3\,,
\end{equation}
\begin{equation}
\beta=\frac{\Lambda}{6\tilde{m}^2[(z_0+1)^3+4\Lambda/3\tilde{m}^2]}e^{\frac{3\tilde{m}^2((z_0+1)^3+
4\Lambda/3\tilde{m}^2)}{\Lambda}}\,.\label{betabeta}
\end{equation}
Here we have used Eqs.~(\ref{eins})-(\ref{vier}). We must stress that, by using our formula, it turns out that $b\sim \exp(-R/\Lambda)\ll (y_0/R_0)^2$, which we have not been able to evaluate, its value being extremely close to zero. The frequency $\mathcal{F}(z_0)$ of dark energy oscillations is
\begin{equation}
\mathcal{F}(z_0)=\sqrt{\frac{\Lambda}{6\tilde{m}^2[(z_0+1)^3+4\Lambda/3\tilde{m}^2]}\,
e^{\frac{3\tilde{m}^2((z_0+1)^3+4\Lambda/3\tilde{m}^2)}{\Lambda}}}\frac{1}{(z_0+1)}\,, \label{betabetabeta}
\end{equation}
while the amplitude decreases as $\exp(3/2(z_0+1))(z-z_0)$. We can verify the validity of our formula by analyzing in detail the graphics of $y_{DE}(z)$ in the vicinity of $z_0=2.60$, $2.55$ and $2.50$. Such graphics are shown in Figs.~5-7. We have chosen an interval of $|z-z_0|=0.02$. The period $T(z_0)$ of dark energy oscillations has to be $T(z_0)=2\pi/\mathcal{F}(z_0)$ and the number of crests in our interval is $n_{|z-z_0|}(z_0)=0.02/T(z_0)$. The predicted values are
\begin{equation*}
T(2.60)\simeq 0.003\,,\phantom{space}n_{0.02}(2.60)\simeq 6.667\,;
\end{equation*}
\begin{equation*}
T(2.55)\simeq 0.004\,,\phantom{space}n_{0.02}(2.55)\simeq 5.000\,;
\end{equation*}
\begin{equation*}
T(2.50)\simeq 0.006\,,\phantom{space}n_{0.02}(2.50)\simeq 3.333\,.
\end{equation*}
These values are in good accordance with the numerical computation. The dark energy density is very close to $\Lambda/3\tilde{m}^2\simeq 2.64333$ and one can check that the amplitude of oscillation decreases with $z$.
\begin{figure}[-!h]
\begin{center}
\includegraphics[angle=0, width=0.55\textwidth]{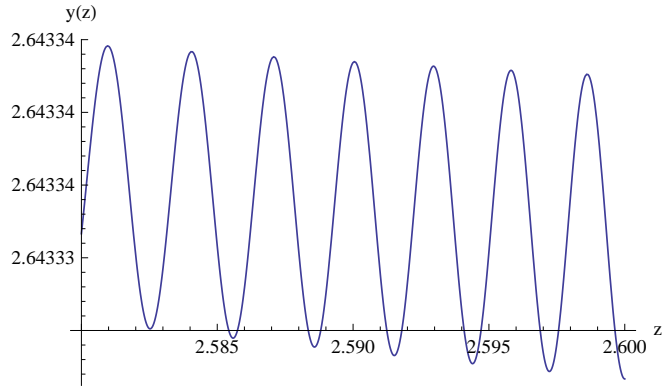}
\end{center}
%  \begin{center}
\caption{Plot of $y_H(z)$ in the vicinity of $z_0=2.60$.\label{Fig5}}
%\end{center}
\end{figure}
\begin{figure}[-!h]
\begin{center}
\includegraphics[angle=0, width=0.55\textwidth]{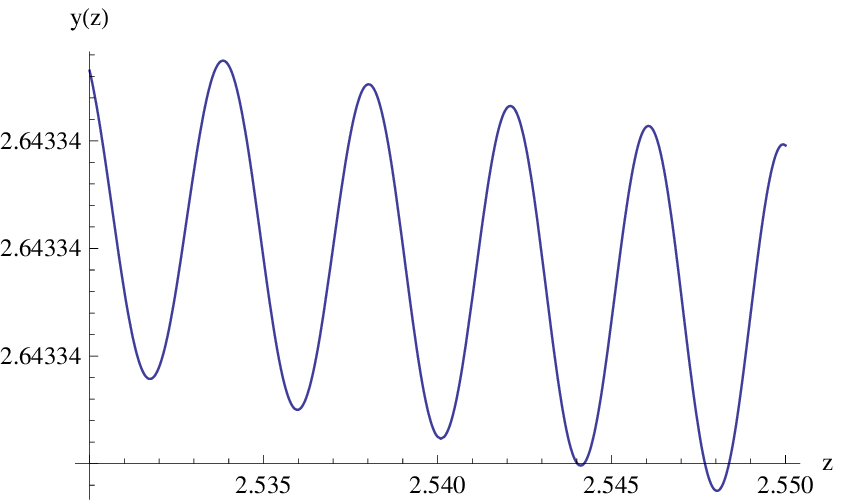}
\end{center}
%  \begin{center}
\caption{Plot of $y_H(z)$ in the vicinity of $z_0=2.55$.\label{Fig6}}
%\end{center}
\end{figure}
\begin{figure}[-!h]
\begin{center}
\includegraphics[angle=0, width=0.55\textwidth]{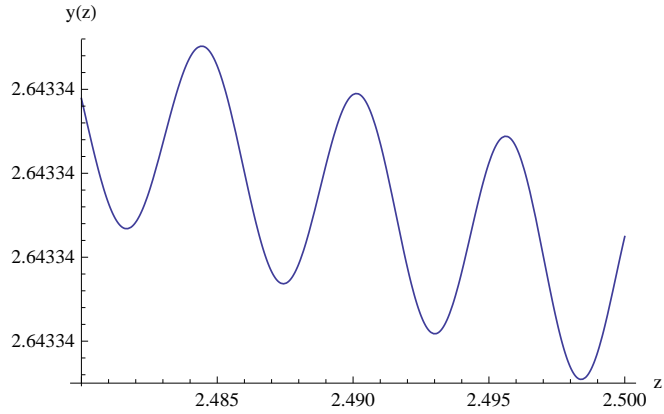}
\end{center}
%  \begin{center}
\caption{Plot of $y_H(z)$ in the vicinity of  $z_0=2.50$.\label{Fig7}}
%\end{center}
\end{figure}

Since the dependence of the amplitude $\mathcal{F}(z_0)$ on $z_0$ is weaker when $z_0<2.5$, we can study the oscillations of the dark energy density around $z_0=2.45$ and $z_0=2.40$ by using a longer interval $|z-z_0|=0.05$. We find
\begin{equation*}
T(2.45)\simeq 0.008\,,\phantom{space}n_{0.05}(2.45)\simeq 6.250\,;
\end{equation*}
\begin{equation*}
T(2.40)\simeq 0.010\,,\phantom{space}n_{0.05}(2.40)\simeq 5.000\,.
\end{equation*}
These results can be compared with Figs.~8-9.
\begin{figure}[-!h]
\begin{center}
\includegraphics[angle=0, width=0.55\textwidth]{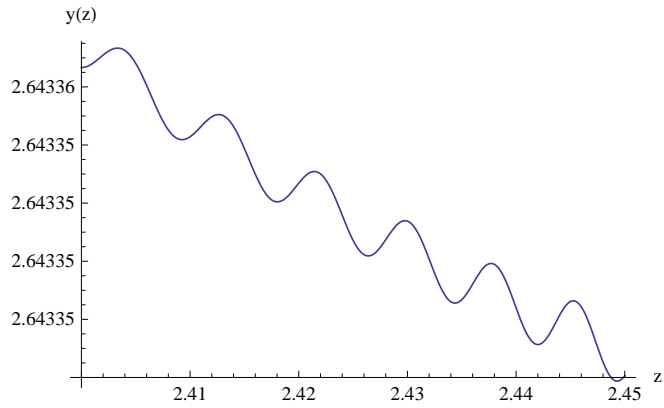}
\end{center}
%  \begin{center}
\caption{Plot of $y_H(z)$ in the vicinity of  $z_0=2.45$.\label{Fig8}}
%\end{center}
\end{figure}
\begin{figure}[-!h]
\begin{center}
\includegraphics[angle=0, width=0.55\textwidth]{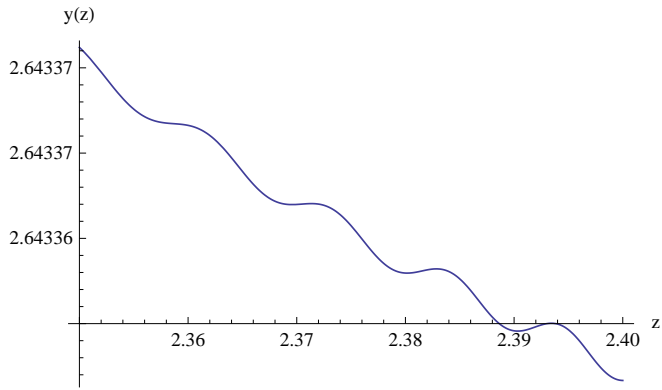}
\end{center}
%  \begin{center}
\caption{Plot of $y_H(z)$ in the vicinity of  $z_0=2.40$.\label{Fig9}}
%\end{center}
\end{figure}

When $z<2.40$ the periods become too large with respect to the change of $\mathcal{F}(z_0)$ and we can no more distinguish oscillations in the dark energy until the beginning of the de Sitter epoch, when the dark energy behavior is governed by Eq.~(\ref{oscillatorysolution}). On the other hand, the effects of such oscillations are amplified in the expression (\ref{omegaoscillatingdue}) for $\omega_{DE}$, where the amplitude of oscillations is proportional to the period. Fig.~10 shows the behavior of $\omega_{DE}$ inside the region $2.3<z<2.6$. Since the frequency of $\omega_{DE}$ is the same as for $y_{DE}(z)$, we can observe there how it decreases with $z$.
\begin{figure}[-!h]
\begin{center}
\includegraphics[angle=0, width=0.55\textwidth]{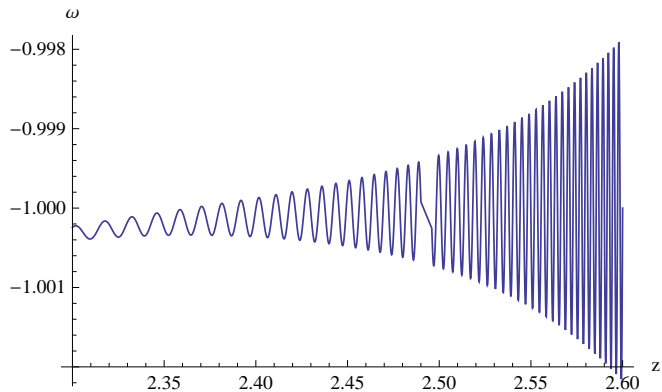}
\end{center}
%  \begin{center}
\caption{Plot of $\omega_{DE}$ as a function of $2.3<z<2.6$.\label{Fig10}}
%\end{center}
\end{figure}

\section{Oscillations and singularities}

%  A finite-time singularity occurs when Hubble parameter or some its derivatives (and therefore energy density, scale factor or, more simplicity, the Riemann tensor components) diverge in a finite time into the future. The softer kinds of such singularities are the Type IV, giving by the Hubble parameter
%  \begin{equation}
%  H=H_0+h_0(t_0-t)^\beta\,,
%  \end{equation}
%  where $\beta$ is a constant so that $\beta>1$. Here, $h_0$, $H_{0}$ and $t_{0}$ are positive constants and the cosmic time $t$ has to be $t<t_{0}$ because it should be for expanding universe. When $t$ is close to $t_{0}$, some high-order derivatives of $H$ and therefore the curvature become singular. We note that in the case of IV Type singularities the Hubble parameter and the Ricci scalar $R$ (which depends on $H$ and $\dot{H}$) do not diverge.

A finite-time, future singularity occurs when the Hubble parameter or some of its derivatives (and, therefore, the energy density, scale factor or, more simply, the Riemann tensor components) diverge \cite{classificationSingularities}. The simplest well-known example of this future singularity
is phantom-induced Big Rip \cite{ckw07}.
The softer kind of such singularities are of Type IV. A Type IV singularity appears if the Hubble parameter and the Ricci scalar $R$ (which depends on $H$ and $\dot{H}$) do not diverge, but higher derivatives of $H$ do tend to infinity. Generally speaking, singularities are called `finite-time' if they occur in a finite time during the cosmological evolution of the universe, yielding nonphysical components of the Riemann Tensor.

Let us return to exponential gravity models. By fitting parameters, they could become indistinguishable with respect to the $\Lambda$CDM model. They mimic, in fact, the cosmological constant to high precision and $\omega_{DE}$ can be made extremely close to the value of $-1$. The transition crossing the phantom divide does not cause any serious problem to the accuracy of the cosmological evolution arising from such kind of models. In Fig.~11 we depict the behavior of the Ricci scalar for the exponential model of the previous chapter. We can see in fact how it decreases with the red-shift, as one would expect, and the amplitude oscillations of the dark energy density are so small that one cannot observe them.
\begin{figure}[-!h]
\begin{center}
\includegraphics[angle=0, width=0.55\textwidth]{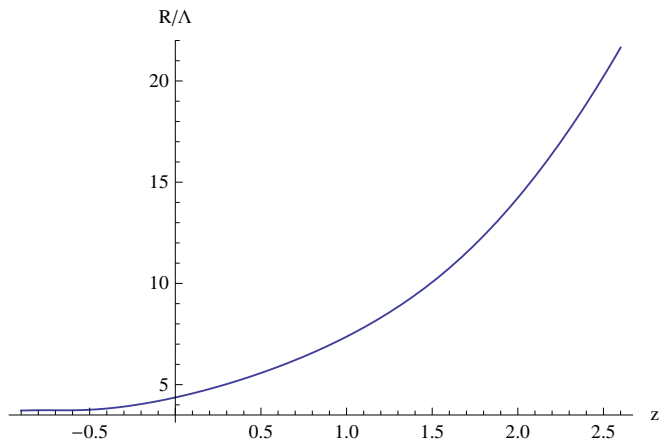}
\end{center}
%  \begin{center}
\caption{Plot of $R/\Lambda$ as a function of $z$.\label{Fig11}}
%\end{center}
\end{figure}

On the other hand, we have seen how, in a model of this kind, $f''(R\rightarrow\infty)\rightarrow 0^+$. As a consequence, the frequency $\mathcal{F}(z_0)$ of the dark energy oscillations increases with the redshift, as in Eq.~(\ref{betabetabeta}). Since the method we have used to analytically evaluate such oscillations yields a linear expansion of the EoM, we cannot observe any singularity in the time derivative of $y_H(z)$ and, correspondingly, in the Hubble parameter. But we have that
\begin{equation}
\left| \frac{d^n}{d t^n}H(t)|_{t_0}\right|\propto \left(\mathcal{F}(z_0)\right)^n\,,
\end{equation}
where $t_0$ is the time at redshift $z_0$. For extremely large values of the redshift, the high time derivatives of the Hubble parameter become infinitely large and approach an effective future singularity, showing a different behavior of the exponential model with respect to the $\Lambda$CDM one. This fact suggests that, back into the past, other terms could have become important in modified gravity theories. For example, in Ref.~\cite{singularities,singF(G)} it is shown how terms proportional to $R^2$ protect the theory against future time singularities and yield a finite value for $f''(R)$.
%The addition of such terms is justified by the necessity to reproduce the early-time acceleration of our universe, namely the inflation epoch.
Simultaneously, these terms induce
also an early-time inflation, thus unifying primordial inflation with dark energy \cite{F(R),no03}

A simple modification of the exponential model which incorporates the
inflationary era is given by a combination of the function discussed
above with another function, which reproduces the cosmological constant during
inflation. A natural possibility is \cite{inflationstep}
\begin{equation}
F(R)=R-2\Lambda\left(1-\mathrm{e}^{-\frac{R}{R_{0}}}\right)
-\Lambda_{i}\left(1-\mathrm{e}^{-\left(\frac{R}{R_i}\right)^n}\right)
+\gamma R^\alpha\,, \label{total}
\end{equation}
where $R_{i}$ and $\Lambda_{i}$ assume the typical values of the curvature
and expected cosmological constant during inflation, namely $R_{i}$,
$\Lambda_i$ $\simeq 10^{20-38} \text{eV}^2$, while $n$ is a natural
number larger than one.
The last term of such equation $\gamma R^\alpha$, where $\gamma$
is a positive dimensional constant and $\alpha$ a real number, is actually
necessary in order to obtain an exit from inflation. Typically,
$\gamma\sim 1/R_i^{\alpha-1}$ and $\alpha>2$, so that the de Sitter solution of inflation results to be unstable and the effects of this term vanish during the matter era, when $R\ll R_i$. The detailed analysis of this extended class of exponential models for early- and late-time acceleration has been carried out in Ref.~\cite{inflationstep}. Here we restrict ourselves to note the effects of the power-like function on the dark energy fluctuations during the matter epoch. Now, Eq.~(\ref{betabetabeta}) reads
\begin{equation}
\mathcal{F}(z_0)=\sqrt{\frac{1}{\frac{2R_0}{\Lambda}e^{-\frac{R_0}{\Lambda}}+
\alpha(\alpha-1)\left(\frac{R_0}{R_i}\right)^{\alpha-1}}}\ \frac{1}{(z_0+1)}\,,
\end{equation}
where $R_0$ is the curvature at redshift $z_0$. Since $\alpha>2$, while the exponential term becomes small on approaching the effective singularity at $R_0\rightarrow\infty$, the power-like term becomes larger, and the frequency of dark energy oscillations does not diverge.

\section{Attractors of exponential models}

The exponential model (\ref{total}) has three solutions (critical points) of the theory in vacuum: an unstable de Sitter solution for $R_{dS}\sim R_i$
(for more detailed evaluation see Ref.~\cite{inflationstep}) describing inflation, the second one, which is stable, at $R_{dS}\simeq 4\Lambda$, and the last one, at $R=0$, corresponds to the Minkowski solution of Special Relativity.
When, in early time, the curvature of the universe is of the same order of $R_i$,
the system gives rise to a de Sitter solution where the universe
expands in an accelerating way but, suddenly, it exits from inflation and tends towards
the minimal attractor at $R_{dS}=4\Lambda$, unless the theory develops a singular solution for $R\rightarrow \infty$. In such case, the model could exit from inflation and move into the
wrong direction, where the curvature would grow up and diverge, and a singularity would appear. It has been demonstrated that exponential gravity is actually free from singularities. However, in the very asymptotic limit, when $R\gg R_i$, the model reads
\begin{equation}
F(R\gg R_i)\simeq \gamma R^{\alpha}\,.\label{approx}
\end{equation}
A class of asymptotic solutions of Eqs.~(\ref{uno})-(\ref{due}) in vacuum, for the model of
Eq.~(\ref{approx}), at the limit $t\rightarrow 0^+$ results
\begin{equation}
H(t)=\frac{H_{0}}{t^{\beta}}\,,
\end{equation}
with
\begin{equation}
R\simeq 12\frac{H_0^2}{t^{2\beta}}.
\end{equation}
Here, $H_{0}$ is a large positive constant and $\beta$ a positive
parameter so that $\beta=1$ or $\beta>1$.

This result shows that in the limit $R\rightarrow+\infty$
the model exhibits a past singularity, which could be identified with
the Big Bang one. However, with a change $t\rightarrow(t_0-t)$, we may obtain a future-time singularity solution.
It is important to stress that this kind of solution is disconnected
from the de Sitter phase of inflation, where the term $R$ is of the
same order of $\gamma R^\alpha$, and is therefore not negligible as in Eq.~(\ref{approx}).
In the very asymptotic limit, the scalaron $F'(R)$ results
\begin{equation}
F'(R)=\gamma\alpha R^{\alpha-1}\,.
\end{equation}
We can also evaluate the potential $V_{\rm eff}$ of Eq.~(\ref{scalaroneeqbis}) through integration of $F''(R)(\partial V_{\rm eff}/\partial F'(R))=\partial V_{\rm eff}/\partial R$. One easily finds
\begin{equation}
V_{\rm eff}(R\gg R_i)=\frac{\gamma^2\alpha(\alpha-1)(2-\alpha)}{3(2\alpha-1)}R^{2\alpha-1}\,.
\end{equation}
We observe that, in order to reach the singularity, the scalaron has to cross over an infinite potential barrier ($V_{\rm eff}(R\rightarrow\infty)\rightarrow\infty$) and go to infinity ($F'(R\rightarrow\infty)\rightarrow\infty$). But clearly this dynamical behavior is forbidden.

We may safely assume that, just after the Big Bang, a Planck epoch takes over where
physics is not described by GR and where quantum gravity effects are
dominant. When the universe exits from the Planck epoch, its
curvature is bound to be the characteristic curvature of inflation
and the de Sitter solution takes over. When inflation ends,
the model moves to the attractor de Sitter point. In this
way, the small curvature regime arises, the first term of Eq.~(\ref{total})
becomes dominant and the physics of the $\Lambda$CDM model ---as we have seen in the preceding sections--- is reproduced.

\section{Discussion \label{SectX}}

It is well-know that an important viable condition in modified gravity is the positivity of the second derivative of the cosmological function ($F''(R)>0$) during the matter dominated era. This condition arises from the stability of matter perturbations in the weak-field regime, taking the curvature to be locally constant. If $F''(R)<0$, such perturbations grow up and, as a consequence, the theory becomes strongly unstable \cite{Faraoni}. In our approach, a detailed analysis of these perturbations has been carried out, by considering the dynamical behavior of dark energy during the matter era.
% Since in late-time matter era is not stable, an expansion of the EOM with respect matter curvature is not allowed, and it is necessary to do some physical assumptions in order to find correct results.
Accurate viable conditions for the stability of the cosmological perturbations during the matter era have been found, namely Eqs.~(\ref{x})-(\ref{xx}), which are in perfect agreement with the constraint $F''(R)>0$. We have also seen how an oscillatory behavior of the dark energy can occur during the matter era. This phenomenon leads to oscillations of $\omega_{DE}$, whose frequency and amplitude decay with the red-shift until the current accelerated epoch, while they substantially grow up as a consequence of perturbations around the de Sitter solution of the field equations.

More specifically, we have investigated the oscillations of dark energy around the phantom divide line, $\omega_{DE}=-1$, both during the matter era and also well in the de Sitter epoch. The analysis of the dark energy during the de Sitter epoch could be carried out by expanding the modified equations of motion around the de Sitter solution. It has been seen that the dark energy density approaches a stable de Sitter point either as a positive power function of $ z+1 $ or through an oscillatory behavior whose amplitude decreases as $(z+1)^{3/2}$. We have also found the time dependence of the dark energy perturbations. The analysis of the dark energy during the matter epoch has been performed by using two different local expansions. For high values of the red shift, $z\gg 3$, the matter epoch is a stable point of the theory and, thus, it is indeed possible to expand the $F(R)$-functions in terms of the dark energy perturbations. On the other hand, in the late-time matter era, $
 1\lesssim z\lesssim 3$, we have considered the realistic case where the dark energy tends to a constant. Then, the results obtained have been accurately confirmed by a numerical computation on a specific model of exponential gravity.

As last point, some problems appearing in relation with the big values generated for the dark energy fluctuations, for large values of the red shift, have been discussed in the paper in detail. When $F''(R)$ is very close to zero, simultaneously with the fact that modified gravity mimics very well the $\Lambda$CDM model, these fluctuations approach an effective singularity in high derivatives of the Hubble parameter, which become infinitely large. This fact suggests that a correction of the Einstein equations in the small curvature region is related with a modification of gravity at high curvature, produced by the leading terms of inflation. Our result can be easily generalized to other different models of dark energy, as the Little Rip one, where the effective $w_{DE}$ approaches the value $-1$ for $t\to \infty$ asymptotically. Note that our approach can be applied to other modified gravities, too (see, for instance, the review \cite{capdl1a}).

We have provided a detailed discussion on the critical points in the matter era. In this set up only the perturbation associated with the scalaron have been investigated, anyhow the results obtained are quite interesting and useful. They have been expressed both in semi-analytical and in numerical form, and we have shown in the paper that they are strictly consistent with each other.

\section*{Acknowledgments}

This research has been supported by the ongoing Bilateral Project INFN(Italy)--MICINN(Spain)
No.~AIC10-D-000560,
by projects FIS2006-02842 and FIS2010-15640, by CPAN Consolider Ingenio Project and
AGAUR 2009SGR-994 (EE and SDO). EE's research was performed in part while on leave
at Department of Physics and Astronomy, Dartmouth College, 6127 Wilder Laboratory,
Hanover, NH 03755, USA.

\end{document}